# Site-ordering/disordering-induced magnetic textures in a vdW ferromagnet by competing global and broken inversion-symmetry

Haoyan Zhang[1,2,+], Jianfeng Guo[1,2,+], Cong Wang[1,2,+], Le Lei[1,2], Shuo Mi[1,2], Songyang Li[1,2], Congkuan Tian[1,2], Shaohua Yan[1,2], Hanxiang Wu[1,2], Shiyu Zhu[3], Rui Xu[1,2], Xueyun Wang[4], Hechang Lei[1,2], Peng Cheng[1,2], Fei Pang[1,2], Wei Ji[1,2], and Zhihai Cheng[1,2,*]

[1]*Key Laboratory of Quantum State Construction and Manipulation (Ministry of Education), Department of Physics, Renmin University of China, Beijing 100872, China*

[2]*Beijing Key Laboratory of Optoelectronic Functional Materials & Micro-nano Devices, Department of Physics, Renmin University of China, Beijing 100872, China*

[3]*Institute of Physics, Chinese Academy of Sciences, Beijing 100190, China*

[4]*School of Aerospace Engineering, Beijing Institute of Technology, Beijing 100081, China*

**Abstract:** $Fe_5GeTe_2$ single crystals can be divided into nonquenched (NQ) and quench-cooled (QC) phases with different magnetic properties. A comprehensive understanding of the magnetic property variations in the NQ and QC phases is imperative for guiding $Fe_5GeTe_2$ towards spintronics applications; however, it remains elusive. Here, we report a real-space study on the structural and magnetic properties of these two magnetic phases using cryogenic magnetic force microscopy and scanning tunneling microscopy. The thermal history introduces disorder and order to the Fe(1) sites, resulting in the NQ and QC phases exhibiting global and broken inversion symmetry, respectively. The observed magnetic domain transitions (branching to labyrinthine) in the spin reorientation process and the distinct 3D spin textures stabilized by magnetic dipolar interaction observed in field-dependent studies allow the NQ phase to exhibit a more resilient global magnetic state. In contrast, the QC phase exhibits enhanced magnetic anisotropy, resulting in a higher $T_C$. Meanwhile, the Dzyaloshinskii-Moriya interaction (DMI) introduced by the broken inversion symmetry causes the QC phase to exhibit a localized magnetic state: no domain transformation occurs during spin reorientation, and irregular domain states are observed in field-related studies. Our work provides an important reference for understanding the complex magnetic properties in $Fe_5GeTe_2$.

+ These authors contributed equally to this work.
* To whom correspondence should be addressed: zhihaicheng@ruc.edu.cn

**Introduction**

Van der Waals (vdW) magnetic materials with quasi-two-dimensional (2D) crystal structures have received extensive research attention in spintronics due to their controllable dimensionality and stacking properties, which can be used to constructing artificial structures with fascinating characteristics.[1-7] However, the majority of reported 2D ferromagnetic (FM) materials, such as $CrI_3$ and $Cr_2Ge_2Te_6$,[8-11] exhibit low Curie temperatures ($T_C$), which limits their future applications in spintronic devices. Recently, the $Fe_nGeTe_2$ family (n = 3, 4, 5) has been reported as a promising vdW ferromagnet due to its high $T_C$ near room temperature (205-330 K) and exotic properties.[12-17] In particular, $Fe_5GeTe_2$, which possesses the highest $T_C$ (275-330 K), has received much attention.[18-36]

A thorough understanding of the physical properties of vdW ferromagnets is essential before to their application in spintronic devices. However, for $Fe_5GeTe_2$, the additional Fe layer not only enhances magnetic interactions to elevate the $T_C$ but also introduces greater complexity into its magnetic properties.[37-44] In particular, the Fe(1) layer, located in the outermost part of the $Fe_5Ge$ sublayer, exhibits a strong correlation with the properties of $Fe_5GeTe_2$.[18,19,42-44] Therefore, the thermal history that can modulate the ordering of the Fe(1) layer exerts a significant influence on the properties of $Fe_5GeTe_2$. For instance, the $T_C$ of quench-cooled (QC phase) crystals is higher compared to that of non-quenching (NQ phase) crystals.[19] The recent discovery reveals that NQ and QC phases $Fe_5GeTe_2$ with different structures exhibit significantly divergent electrical properties.[31] Topological node lines are observed in the NQ phase with a disordered Fe(1) layer, while flat bands are observed in the QC phase with an ordered Fe(1) layer. Considering the coupling between structure and magnetism, this symmetry breaking caused by the different structure of the Fe(1) layer should also have a significant effect on the magnetic properties of $Fe_5GeTe_2$. However, detailed comparative studies regarding the magnetic properties of these two phases remain relatively scarce, especially at the microscopic scale.

In this work, we have systematically characterized the structure and magnetic properties of NQ and QC phases of $Fe_5GeTe_2$ using cryogenic magnetic force microscopy (MFM) and scanning tunneling microscopy (STM). For the NQ phase $Fe_5GeTe_2$, the global inversion symmetry due to disorder at the Fe(1) sites allows it to exhibit a global magnetic state that is intrinsically elastic and scalable. Therefore, the NQ phase undergoes a transformation by the magnetic domain (branching into labyrinthine domains) simultaneously with the spin reorientation transition. Further field-dependent studies have revealed distinct 3D topological spin textures stabilized in the NQ phase by magnetic dipolar interactions, including skyrmion cocoons and a skyrmion lattice. For the QC phase $Fe_5GeTe_2$, the STM topography reveals a $\sqrt{3}\times\sqrt{3}$ superstructures formed due to the ordering of the Fe(1) sites. The broken inversion symmetry induces the Dzyaloshinskii-Moriya interaction (DMI), leading to the manifestation of a localized magnetic state in the QC phase. Therefore, the QC phase does not exhibit magnetic domain (flower-like domain) transitions during the spin reorientation process and irregular magnetic domain states are observed in further field correlation studies. In addition, the ordering of Fe(1) sites enhances the FM exchange coupling and magnetic anisotropy, thereby resulting in a higher $T_C$ for the QC phase. Our results not only provide essential guidance for the future implementation of $Fe_5GeTe_2$ in spintronic devices, but also serve as a significant complement to the physical properties of vdW ferromagnets.

# Results

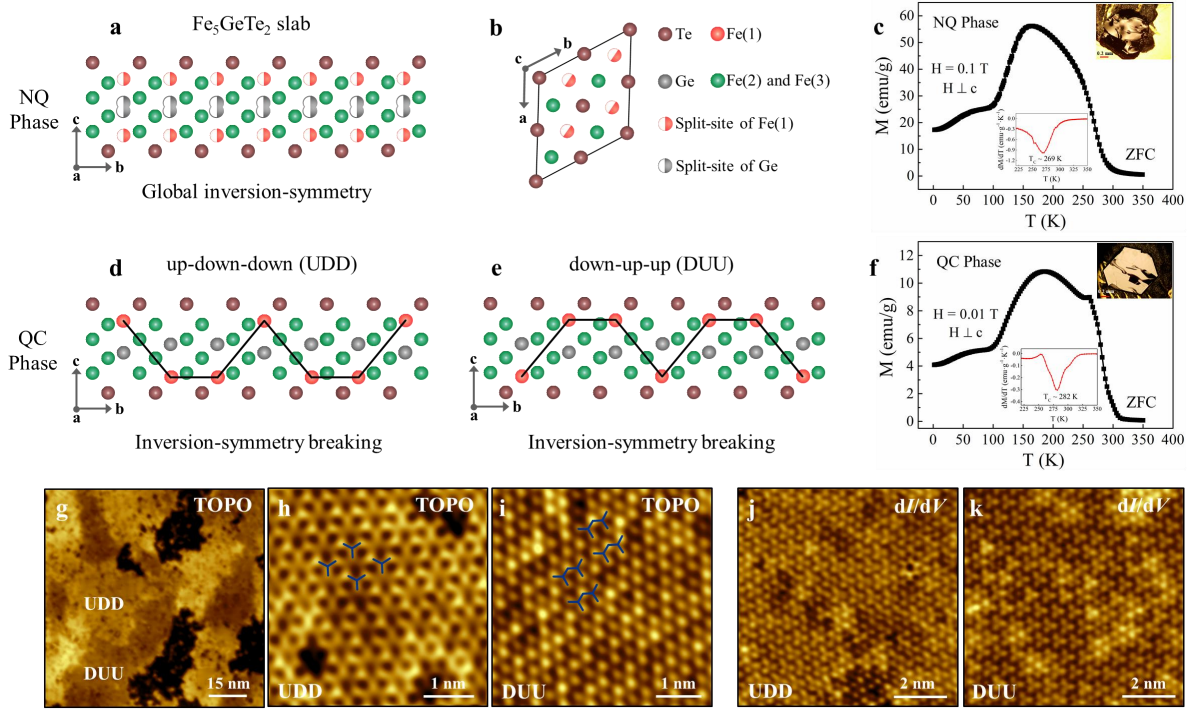

**Figure 1. Crystal structure and magnetic characterizations of Fe$_5$GeTe$_2$ single crystal. (a,b)** Schematic illustration of the crystal structure of Fe$_5$GeTe$_2$. (a) and (b) correspond to the side view and top view, respectively. Fe(1) and Ge were modeled as split sites. **(c)** Temperature dependence of zero field cooled (ZFC) magnetic susceptibility (H⊥c) of NQ phase Fe$_5$GeTe$_2$. The lower inset shows the differential curve corresponding to the M-T curve. **(d,e)** Two equivalent Fe(1) site-ordered $\sqrt{3}\times\sqrt{3}$ superstructures in the QC phase: up-down-down (UDD) ordering (d) and down-up-up (DUU) ordering (e). **(f)** Temperature dependence of ZFC magnetic susceptibility of QC phase Fe$_5$GeTe$_2$. **(g)** Large-scale STM topographic image of QC phase Fe$_5$GeTe$_2$, in which the UDD and DUU regions are marked respectively. **(h-k)** High-resolution STM images (h,i) and $dI/dV$ conductance maps (j,k) of the visible $\sqrt{3} \times \sqrt{3}$ superstructures of the UDD and DUU region in the QC phase. Imaging parameters: $V_b$ = -0.7 V, $I_t$ = -30 pA (g); $V_b$ = -70 mV, $I_t$ = -130 pA (h); $V_b$ = -20 mV, $I_t$ = -150 pA (i); $V_b$ = -50 mV, $I_t$ = -100 pA (j,k).

Fe$_5$GeTe$_2$ is a member of the vdW ferromagnet Fe$_n$GeTe$_2$ family (n = 3, 4, 5), in which each layer consists of a Fe$_5$Ge layer sandwiched by Te layers, as shown in Figure 1a,b. The crystal structure of Fe$_5$GeTe$_2$ corresponds to the rhombohedral $R\bar{3}m$ space group, which exhibits centrosymmetry with lattice parameters a = b ≈ 4.04 Å and c ≈ 29.19 Å.[18] There are three inequivalent Fe sites in Fe$_5$GeTe$_2$, namely Fe(1), Fe(2), and Fe(3). Among these sites, both the Fe(2) and Fe(3) sites are fully occupied, while Fe(1) is located in the outermost Fe$_5$Ge sublayer with two possible split-sites. Affected by the Fe(1) site, there are also two split-sites for Ge, which tend to occupy sites away from the Fe(1). The selection of Fe(1) sites in space is not entirely independent, and the thermal history can dictate the ordering and disordering of Fe(1) sites.[19,31] According to different thermal histories (Figure S1), $Fe_5GeTe_2$ can be divided into NQ phase and QC phase single crystals with different structural symmetries. The X-ray diffraction patterns were able to demonstrate the high quality of the single crystals used in our experiments (Figure S2).

The Fe(1) sites of crystals obtained by slow cooling (NQ phase) tend to be randomly distributed with global inversion symmetry, as shown in Figure 1a. In contrast, the Fe(1) sites of crystals obtained by quenching (QC phase) form a UDD or DUU pattern with broken inversion symmetry, as shown in Figure 1d,e. To further confirm this, we conducted STM measurements on QC phase crystals. The large-scale STM topographic image reveals the presence of structural domains with UDD and DUU configurations, as shown in Figure 1g. The atomic resolution images of UDD and DUU regions are shown in Figure 1h,i. Due to the ordering of Fe(1) sites, a $\sqrt{3}\times\sqrt{3}$ superstructure is formed by the topmost Te layer, consistent with previous reports.[43,44] Furthermore, the ordering of Fe(1) sites also influences the electron distribution, leading to the observation of a $\sqrt{3}\times\sqrt{3}$ superstructure in the *dI/dV* conductance maps of both UDD and DUU regions, as shown in Figure 1j,k. According to our STM results, it can be concluded that the QC phase $Fe_5GeTe_2$ indeed exhibits a broken inversion symmetry. It is worth noting that the broken inversion symmetry introduces additional DMI, while the regions with UDD and DUU configurations exhibit DMI in opposite directions (Figure S3). Therefore, the presence of fine structural domains is expected to influence the magnetic state of the QC phase.

The difference in structural symmetry results in significant variations in the magnetic properties between the NQ phase and QC phase $Fe_5GeTe_2$. Figure 1c,f shows the temperature dependence of magnetization (M-T) curve for the NQ and QC phases under ZFC, with a magnetic field applied along *ab* plane. In order to obtain more accurate $T_C$ for the two phases, we determine $T_C$ by identifying the differential minimum in the M-T curve. Thus, the $T_C$ of the NQ and QC phases are 269 K and 282 K, respectively. The ordering of Fe(1) sites results in a higher $T_C$ for the QC phase $Fe_5GeTe_2$. The macroscopic magnetic transport difference implies a substantial divergence in the microscopic magnetic properties between the NQ and QC phases. To obtain a comprehensive understanding of the magnetic property differences resulting from symmetry breaking between these two phases at the microscopic scale, we conducted real-space MFM characterization.

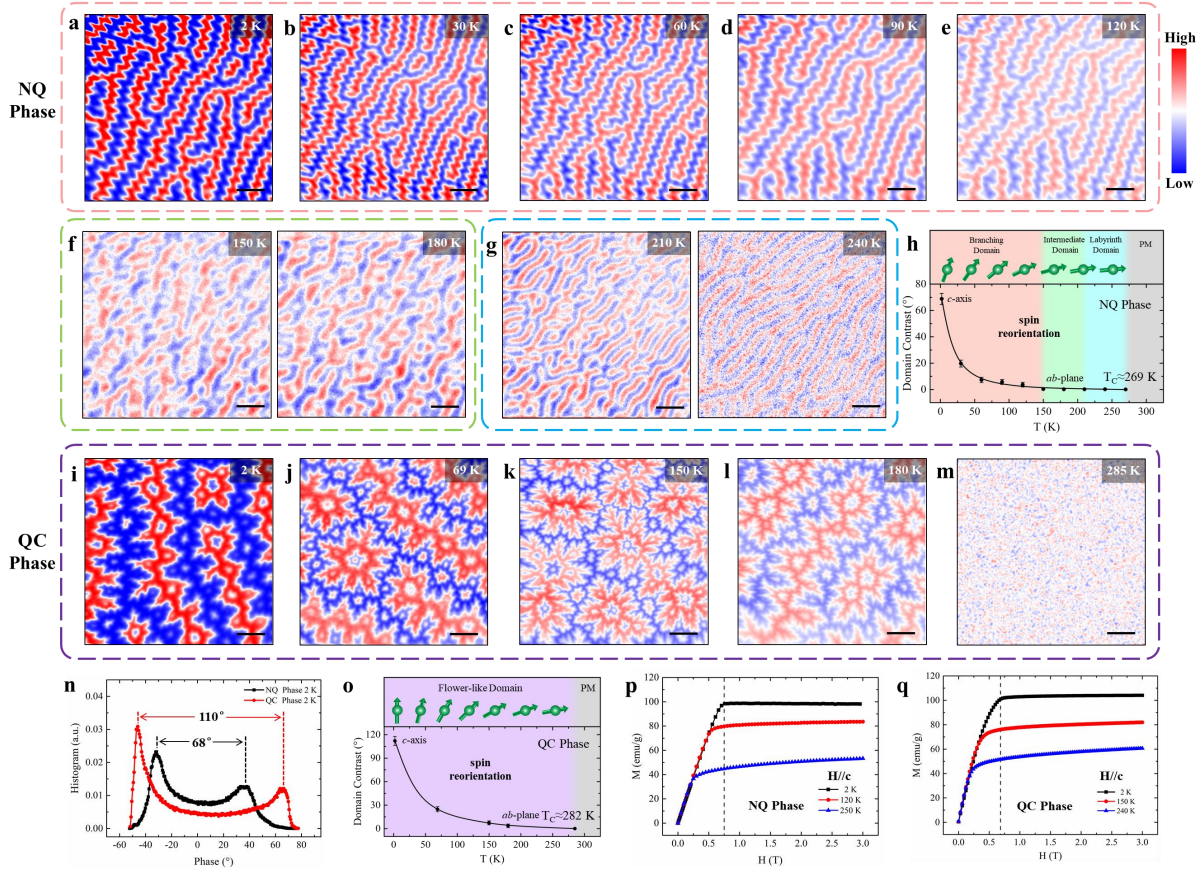

**Figure 2. MFM characterizations of NQ and QC phases Fe$_5$GeTe$_2$ at variable temperatures. (a-g)** MFM images of NQ phase with increasing temperature. The temperature-dependent spin-reorientation process can be divided into three stages: branching domain (a-e), intermediate domain (f), and labyrinth domain (g). **(h)** Magnetic domain contrast of NQ phase versus the temperature with the schematic spin-reorientation illustration, showing the temperature-dependent magnetic states. **(i-m)** MFM images of QC phase with increasing temperature. **(n)** Histogram of the MFM signals for NQ and QC phases at 2 K. **(o)** Magnetic domain contrast of QC phase versus the temperature with schematic spin-reorientation illustration, showing stronger out-of-plane magnetic anisotropy. **(p,q)** Field-dependent magnetizations at different temperatures with H//c for NQ (p) and QC (q) phases. Scale bar: 3 μm.

Figures 2a-g show typical MFM images of NQ phase Fe$_5$GeTe$_2$ with increasing temperature (2-270 K). The MFM image at 2 K (Figure 2a) exhibits a typical branching FM domain with a contrast of 68° between different domains (Figure S4). The MFM signal is proportional to the out-of-plane stray field gradient, indicating that the easy magnetization axis of NQ phase Fe$_5$GeTe$_2$ at 2 K predominantly aligns with the *c*-axis. With the increase in temperature, there is no significant change observed in the magnetic domain (Figures 2b-e),

while the contrast of domains rapidly decreases (Figure S4). This suggests the occurrence of a spin reorientation transition, wherein the easy magnetization axis shifts from the *c*-axis to the *ab* plane.[27] The same transition can also be obtained from the M-H curve (Figure S5). When the temperature reaches 150 K, the magnetic moment turns to *ab* plane and domain contrast decreases to less than 0.3° (Figure S4). Simultaneously, the magnetic domains enter an intermediate chaotic state (Figure 2f). By 210 K, the magnetic domains gradually transform into labyrinth domains (Figure 2g). Subsequently, this domain remains stable until the crystal transitions into a paramagnetic state (270 K). The domain contrast of the whole warming process was extracted and plotted as a curve, as shown in Figure 2h. The entire process can be divided into three distinct stages: branching domain (2-150 K), intermediate domain (150-210 K), and labyrinth domain (210-269 K).

Figures 2i-m show typical MFM images of QC phase $Fe_5GeTe_2$ with increasing temperature (2-285 K). Unlike the NQ phase, the MFM image of the QC phase at 2 K exhibits flower-like FM domains (Figure 2i). The contrast in different domains is 110°, which is larger than that of the NQ phase, as shown in Figure 2n, indicating a stronger perpendicular magnetic anisotropy (PMA) for the QC phase $Fe_5GeTe_2$. This likely accounts for its high $T_C$. The spin reorientation transition is also observed with increasing temperature, as shown in Figure 2o; however, there is no significant change in the magnetic domains of the QC phase (Figure 2j-m). In addition, the enhanced magnetic anisotropy results in a higher domain contrast of the QC phase compared to that of the NQ phase at the same temperature (Figure S6). The M-H curves for NQ and QC phases are shown in Figure 2p,q, respectively. The QC phase exhibits a lower saturation field and a higher saturation magnetization.

The magnetic behavior difference of $Fe_5GeTe_2$ originates from the Fe(1) sites. The global inversion symmetry resulting from disorder at the Fe(1) sites leads to the manifestation of a global magnetic state in the NQ phase. The absence of distinct structural domains makes it easier for the magnetic moments in the NQ phase to evolve with external conditions. Consequently, during the spin reorientation transition, the magnetic domains adjust to accommodate the change in the easy magnetization axis. In comparison, the Fe(1) sites of the QC phase have a UDD or DUU configuration with broken inversion symmetry, which makes the QC phase behave as a localized magnetic state. The presence of structural domains coupled with their introduction of stray DMI exerts a pinning effect on the magnetic domains in the QC phase. As a result, the magnetic domains do not change significantly during the spin reorientation transition. Moreover, the ordering of Fe(1) sites effectively enhances the FM exchange coupling and magnetic anisotropy in the QC phase,[42] thereby resulting in an increase in the $T_C$. In the following, additional field-dependent measurements are employed to further elucidate the difference between the two magnetic states.

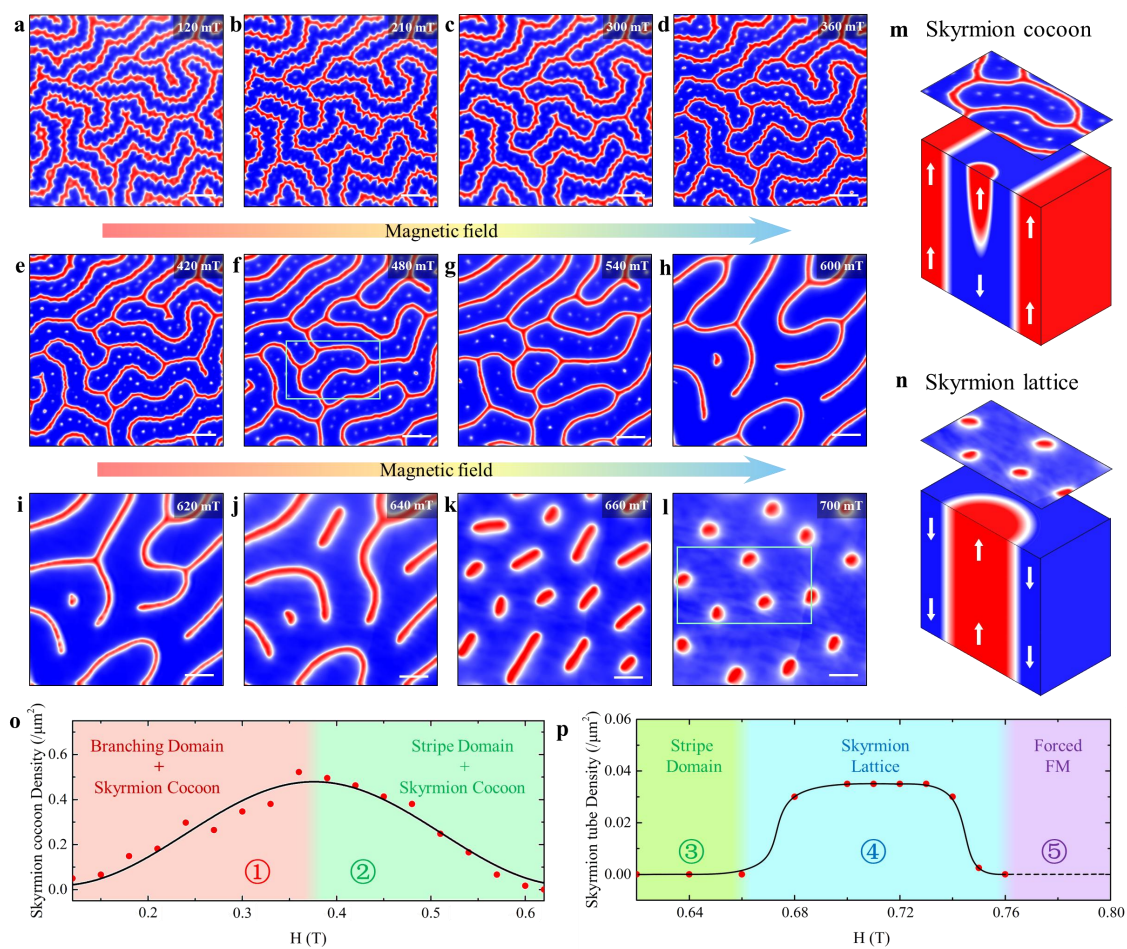

**Figure 3. Various spin textures of NQ phase Fe$_5$GeTe$_2$ in field-dependent MFM characterization. (a-l)** MFM images of NQ phase with increasing applied magnetic fields (H//*c*) at 2 K. The magnetic field is parallel to the *c* axis and along the -*c* direction. The magnetic field-dependent spin textures evolution can be divided into three stages: increasing skyrmion cocoon (a-d), decreasing skyrmion cocoon (e-h), stripe domain and skyrmion lattice (i-l). **(m,n)** Schematic representation of the emergent skyrmion cocoons (m) in (a-h) and the skyrmion littice (n) evolved from the stripes in (i-l). **(o,p)** Skyrmion cocoon (o) and skyrmion tube (p) density versus the applied magnetic field. Scale bar: 3 μm.

The field-dependent MFM images of NQ phase Fe$_5$GeTe$_2$ reveal a variety of 3D topological spin textures, including the stabilization of skyrmionic cocoons and a skyrmion lattice, due to the inherent elasticity and scalability of the global magnetic state. Figures 3a-l show typical MFM images of NQ phase Fe$_5$GeTe$_2$ at 2 K with increasing magnetic field (Figure S9). The whole field-dependent process can be divided into three stages. In stage I (Figure 3a-d), The branching domain aligned with the magnetic field gradually expands, while skyrmionic cocoons emerge within it and progressively increase. The schematic of the skyrmionic cocoon is shown in Figure 3m, exhibiting a relatively weaker MFM signal (Figure S10), which is consistent with previous reports.[45] In stage II (Figure 3e-h), the branching domains transform into stripe domains, while the skyrmionic cocoons are unable to maintain the spin texture and gradually disappears due to the increased strength of the Zeeman effect. In stage III (Figure 3i-l), the stripe domains gradually shorten and shrink to form a skyrmion lattice, schematically shown in Figure 3n. Each individual point within this lattice corresponds to a distinct skyrmion tube that exhibits a relatively strong MFM signal (Figure S10). The density of skyrmion cocoon and skyrmion tube in each MFM image was extracted and plotted as shown in Figures 3o,p. Five different magnetic domain structures are distinguished. The schematic evolution of the magnetic domain structure for the whole field-dependent process is shown in Figure S11. Similar results can be obtained for different variable magnetic field processes (Figures S7,8).

The magnetic field sensitivity of the global magnetic state enables the realization of local magnetic domain manipulation through the utilization of a magnetic tip (Figure S12). The stray field of the magnetic tip can modulate both skyrmionic cocoons and stripe domains. Specifically, the movement and disappearance of the skyrmionic cocoon can be modulated, while also having the ability to regulate the breakage of the stripe domains. In particular, it is possible to realize the cutting of the stripe domains at specific locations. The manipulability of magnetic domains further demonstrates the potential of NQ phase Fe$_5$GeTe$_2$ for applications in spintronics devices.

The subsequent discussion will focus on the factors that contribute to stabilizing 3D topological spin textures in the NQ phase. The generation of stable magnetic skyrmions can be achieved through three primary mechanisms, namely bulk DMI, interfacial DMI, and magnetic dipolar interaction.[46-48] The exclusion of bulk DMI can be deduced from the presence of global inversion symmetry in NQ phase $Fe_5GeTe_2$. Previous reports have shown no significant interfacial DMI in $Fe_5GeTe_2$ as well.[39] Therefore, the diverse spin textures observed in $Fe_5GeTe_2$ are likely attributed to dipolar interactions. The skyrmions generated by dipolar interactions are characterized by a size-magnetic field coupling,[46] which is consistent with our observations (Figure S9) and further validates our hypothesis. For the NQ phase, the periodically arranged magnetic skyrmions arise from the competition between the magnetocrystalline anisotropy and dipolar interactions. In contrast, it is anticipated that the QC phase will exhibit a more intricate spin texture in its field-dependent MFM measurements upon the introduction of DMI.

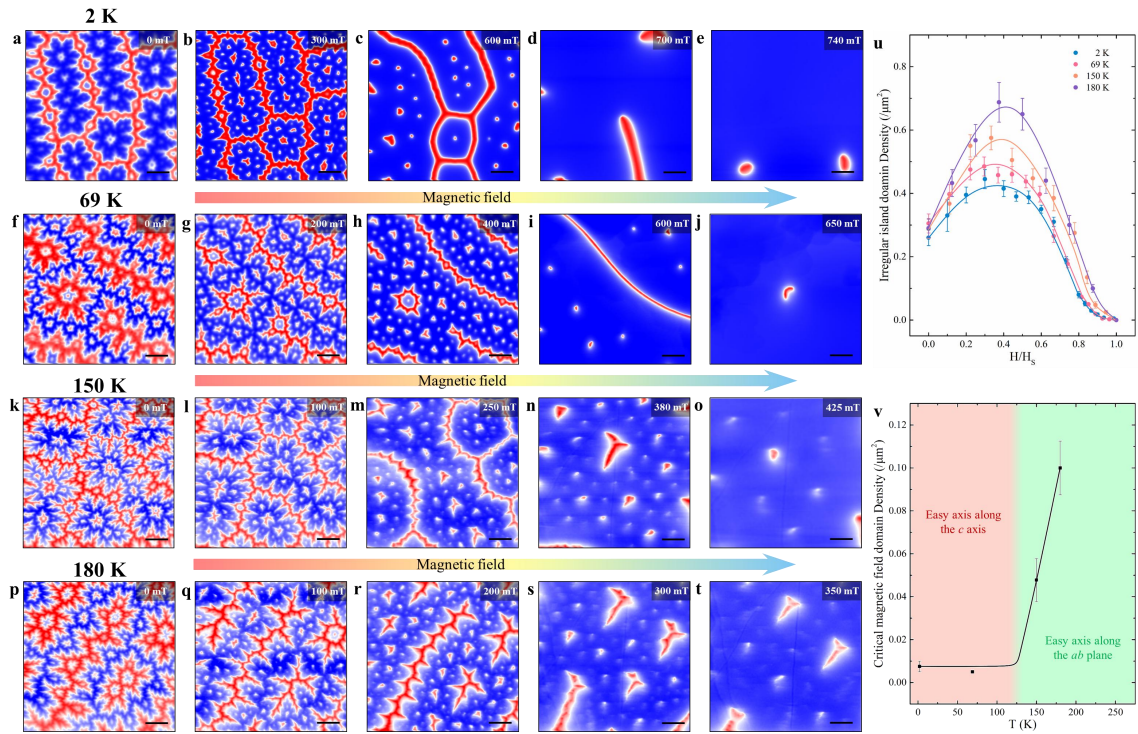

**Figure 4. Evolved spin textures of QC phase $Fe_5GeTe_2$ in field-dependent MFM characterizations. (a-t)** MFM images of QC phase with increasing magnetic fields at 2 K (a-e), 69 K (f-j), 150 K (k-o), and 180 K (p-t). **(u)** Irregular island magnetic domain density at different temperatures versus the normalized magnetic field. **(v)** Magnetic domain density under critical magnetic field versus the temperature. The increase in magnetic domain density indicates a reorientation of the easy magnetization axis from the *c*-axis to the *ab*-plane. Scale bar: 3 μm.

Figures 4a-e show typical MFM images of QC phase $Fe_5GeTe_2$ at 2 K with increasing magnetic field (Figure S13). A stronger PMA results in an enlargement of the magnetic domains within the QC phase compared to the NQ phase. However, due to the presence of stray DMI induced by structural domains, the spins are compelled to align in-plane.

Consequently, in order to achieve equilibrium, additional magnetic domains need to be introduced, ultimately leading to a fragmented domain state resembling a flower-like pattern observed in Figure 4a. Since the structural domains are smaller than the magnetic domains, different orientations of DMI result in a complex magnetic texture in the QC phase. Therefore, field-dependent measurements reveal an irregular magnetic domain state for the QC phase as shown in Figures 4b-d. The coexistence of merons with skyrmions in the QC phase, as recently observed, further substantiates our findings.[37] The skyrmion lattice appears near the saturation magnetic field, as shown in Figure 4e. The absence of observed size-magnetic field coupling suggests that the spin texture of the QC phase stabilized by DMI, unlike the NQ phase. In addition, the magnetic structure of the QC phase is much more stable and no manipulation of the localized magnetic domains is realized.

Representative MFM images of QC phase $Fe_5GeTe_2$ at 69 K (Figure S14), 150 K (Figure S15), and 180 K (Figure S16) for increasing magnetic fields are shown in Figures 4f-t. The field-dependent magnetic domain evolution of the QC phase at different temperatures is relatively similar, but the saturation field gradually decreases. To investigate the impact of temperature on the magnetic state of the QC phase more accurately, we extracted the variation curves of irregular island magnetic domain density with the normalized magnetic field at different temperatures from the MFM images, as shown in Figure 4u. The density of the irregular magnetic domains in the QC phase appears to increase with the temperature, which is attributed to the gradual shift of the easy magnetization axis towards the in-plane direction. The variation in the orientation of the easy magnetization axis, coupled with the presence of stray DMI, leads to more complex magnetic textures within the QC phase. This is also why distinct spin textures can only be observed in the QC phase at higher temperatures.[24,29,37,39] The critical magnetic field (just below the saturation field) domain densities at different temperatures are shown in Figure 4v, which can more intuitively reflect the shift of the easy magnetization axis. In contrast, the NQ phase in the intermediate domain stage (150 K) exhibits complete domain pinning (Figure S17), which can be attributed to the chaotic spin orientation of the intermediate state. The magnetic domains exhibited negligible evolution with increasing magnetic fields, ultimately reaching saturation at 250 mT.

**Discussion**

According to the experimental results, the properties of $Fe_5GeTe_2$ are significantly influenced by symmetry breaking resulting from ordering or disorder at the Fe(1) sites. Electrically, the NQ phase characterized by global inversion symmetry exhibits topological nodal lines, whereas the QC phase featuring broken inversion symmetry manifests flat bands.[15,31] Magnetically, the NQ phase exhibits a resilient and scalable global magnetic state with various 3D spin textures, while the QC phase exhibits a localized magnetic state with stronger PMA, resulting in an increase in $T_C$. Therefore, the modulation of $Fe_5GeTe_2$ properties by thermal history can be utilized to fabricate reversible non-volatile electronic devices. This can provide a new idea in the field of information storage. Additionally, given the intricate magnetic properties of $Fe_5GeTe_2$, there are many questions awaiting further investigation. For instance, the site-ordering of $Fe_5GeTe_2$ crystals obtained by quenching is counterintuitive, highlighting the importance of observing the crystal structure of $Fe_5GeTe_2$ at high temperatures to comprehend its intricate physical properties. Moreover, a more comprehensive investigation is required to determine whether the defects introduced by quenching result in Kondo screening similar to $Fe_3GeTe_2$ for the QC phase $Fe_5GeTe_2$.[17]

Furthermore, our experimental results show that the magnetic properties of the QC phase $Fe_5GeTe_2$ are significantly influenced by its structural domains. In this study, due to the small size of the structural domains, the stray DMI makes the QC phase exhibit fragmented magnetic domains. However, If the size of the structural domains is further increased, the stray DMI will gradually becomes more uniform, resulting in the generation of a stable chiral spin texture in the QC phase. Therefore, the modulation of the structural domains of the QC phase is crucial for regulating its magnetic state. Moreover, given the opposite DMI in UDD and DUU regions, the control of the structural domain could potentially lead to the generation of spin textures exhibiting contrasting chirality in these two distinct structural domains. According to the above results, the subsequent study can attempt to explore the regulation of structural domains by introducing stresses, patterning, etc.

Finally, in order to provide further evidence that the QC phase $Fe_5GeTe_2$ indeed corresponds to a high-temperature phase, we conducted additional long-term evolutionary measurements (Figure S18). The QC phase Fe$_5$GeTe$_2$ is placed in an ultra-high vacuum STM chamber to prevent oxidation. After a few months, the STM image reveals the emergence of disorder in the UDD and DUU regions, suggesting a propensity for transformation from the QC phase Fe$_5$GeTe$_2$ to the NQ phase, which corresponds to a transition from its high-temperature state to a low-temperature state. Simultaneously, it also substantiates the robustness of the site order/disorder induced by thermal history, rendering its transformation under natural conditions arduous.

**Conclusions**

In summary, our MFM results provide insights into the real-space variations in magnetic properties between NQ and QC phases of Fe$_5$GeTe$_2$ arising from distinct thermal histories. The difference arises from the breaking of symmetry caused by the ordering or disorder of the Fe(1) sites. The STM results reveal the emergence of a $\sqrt{3}\times\sqrt{3}$ superstructure through Fe(1) ordering in the QC phase. The ordering of the Fe(1) sites enhances the PMA of QC phase, allowing it to possess a higher T$_C$ than the NQ phase. The temperature-dependent studies reveal that the magnetic domains of the NQ phase undergo a transition from branching domains to labyrinthine domains during the spin reorientation process, while there is no significant change observed in the domains of the QC phase during this transition. Further investigations on field-dependence have revealed the presence of different 3D spin textures stabilized by magnetic dipolar interactions in the NQ phase, while the magnetic domains in the QC phase predominantly exhibit an irregular state due to the stray DMI introduced by the structural domains. Therefore, the NQ and QC phases correspond to a global magnetic state and a localized magnetic state, respectively. Our findings not only contribute significantly to the comprehension of the intricate magnetic properties of Fe$_5$GeTe$_2$, but also have the potential to advance the utilization of Fe$_5$GeTe$_2$ in future spintronic devices.

**Acknowledgments**

This project was supported by the Ministry of Science and Technology (MOST) of China (No. 2023YFA1406500), Strategic Priority Research Program (Chinese Academy of Sciences, CAS) (No. XDB30000000), National Natural Science Foundation of China (NSFC) (No. 61674045 and 61911540074), and Fundamental Research Funds for the Central Universities and Research Funds of Renmin University of China (No. 21XNLG27 and 22XNH097).

## Materials and Methods

**Sample preparation.**

Single crystals of $Fe_5GeTe_2$ were grown through the chemical vapor transport method with iodine as the transport agent.[20,24,40] The NQ phase $Fe_5GeTe_2$ single crystals are obtained by natural cooling from 760 °C to room temperature, while the QC phase $Fe_5GeTe_2$ single crystals are obtained by direct quenching from 760 °C. $Fe_5GeTe_2$ single crystals were cleaved in ambient conditions to expose fresh surfaces before MFM measurements were performed.

**STM measurement.**

The STM experiments were performed with a commercial variable-temperature STM (PanScan Freedom, RHK). The $Fe_5GeTe_2$ single crystals used in experiments were cleaved in ultra-high vacuum chamber at room temperature, and then immediately transferred to STM head. The low temperature STM measurements were performed at 9K with chemical etched W tip, which was calibrated carefully on a clean Ag(111) surface.

**MFM measurement.**

The MFM experiments were conducted using a commercial magnetic force microscope (attoAFM I, attocube) employing a commercial magnetic tip (Nanosensors, PPP-MFMR, Quality factor approximately 1600 at 2 K) based on a closed-cycle He cryostat (attoDRY2100, attocube). The scanning probe system was operated at the resonance frequency, approximately 72 kHz, of the magnetic tip. The MFM images were captured in a constant height mode with the scanning plane nominally ~100 nm (or 200 nm) above the sample surface. The MFM signal, i.e., the change in the cantilever phase, was proportional to the out-of-plane stray field gradient. The dark (bright) regions in the MFM images represented attractive (repulsive) magnetization, where the magnetization was parallel (antiparallel) to the magnetic tip moments.

**XRD and transport measurements.**

The single crystal x-ray diffraction patterns were collected from a Bruker D8 Advance x-ray diffractometer using Cu $K_α$ radiation. The measurements of magnetic property were performed on a Quantum Design magnetic property measurement system (MPMS-3).

**Reference:**

1. K. F. Mak, J. Shan & D. C. Ralph, Probing and controlling magnetic states in 2D layered magnetic materials. *Nat. Rev. Phys.* **1**, 646–661 (2019).
2. C. Gong & X. Zhang, Two-dimensional magnetic crystals and emergent heterostructure devices. *Science* **363**, eaav4450 (2019).
3. B. Huang, M. A. McGuire, A. F. May, et al. Emergent phenomena and proximity effects in two-dimensional magnets and heterostructures. *Nat. Mater.* **19**, 1276–1289 (2020).
4. S. Yang, T. Zhang & C. Jiang, van der Waals Magnets: Material Family, Detection and Modulation of Magnetism, and Perspective in Spintronics. *Adv. Sci.* **8**, 2002488 (2021).
5. W. Zhang, P. K. J. Wong, R. Zhu, et al. Van der Waals magnets: Wonder building blocks for two-dimensional spintronics? *InfoMat.* **1**, 479–495 (2019).
6. K. S. Burch, D. Mandrus & J. G. Park, Magnetism in two-dimensional van der Waals materials. *Nature* **563**, 47–52 (2018).
7. L. Du, T. Hasan, A. Castellanos-Gomez, et al. Engineering symmetry breaking in 2D layered materials. *Nat. Rev. Phys.* **3**, 193–206 (2021).
8. B. Huang, G. Clark, E. Navarro-Moratalla, et al. Layer-dependent ferromagnetism in a van der Waals crystal down to the monolayer limit. *Nature* **546**, 270–273 (2017).
9. Z. Sun, Y. Yi, T. Song, T. et al. Giant nonreciprocal second-harmonic generation from antiferromagnetic bilayer $CrI_3$. *Nature* **572**, 497–501 (2019).
10. C. Gong, L. Li, Z. Li, et al. Discovery of intrinsic ferromagnetism in two-dimensional van der Waals crystals. *Nature* **546**, 265–269 (2017).
11. I. A. Verzhbitskiy, H. Kurebayashi, H. Cheng, et al. Controlling the magnetic anisotropy in $Cr_2Ge_2Te_6$ by electrostatic gating. *Nat. Electron.* **3**, 460–465 (2020).
12. Z. Fei, B. Huang, P. Malinowski, et al. Two-dimensional itinerant ferromagnetism in atomically thin $Fe_3GeTe_2$. *Nat. Mater.* **17**, 778–782 (2018).
13. Y. Deng, Y. Yu, Y. Song, et al. Gate-tunable room-temperature ferromagnetism in two-dimensional $Fe_3GeTe_2$. *Nature* **563**, 94–99 (2018).
14. J. Seo, D. Y. Kim, E. S. An, et al. Nearly room temperature ferromagnetism in a magnetic metal-rich van der Waals metal. *Sci. Adv.* **6**, eaay8912 (2020).
15. F. Wang & H. Zhang. Flat bands and magnetism in $Fe_4GeTe_2$ and $Fe_5GeTe_2$ due to bipartite crystal lattices. *Phys. Rev. B* **108**, 195140 (2023).
16. X. Yang, X. Zhou, W. Feng, et al. Strong magneto-optical effect and anomalous transport in the two-dimensional van der Waals magnets $Fe_nGeTe_2$ (n = 3, 4, 5), *Phys. Rev. B* **104**, 104427 (2021).
17. M. Zhao, B.-B. Chen, Y. Xi, et al. Kondo Holes in the Two-Dimensional Itinerant Ising Ferromagnet $Fe_3GeTe_2$. *Nano Lett.* **21**, 6117-6123 (2021).
18. A. F. May, D. Ovchinnikov, Q. Zheng, et al. Ferromagnetism near Room Temperature in the Cleavable van der Waals Crystal $Fe_5GeTe_2$, *ACS Nano* **13**, 4436 (2019).
19. A. F. May, C. A. Bridges & M. A. McGuire, Physical properties and thermal stability of $Fe_{5-x}GeTe_2$ single crystals. *Phys. Rev. Mater.* **3**, 104401 (2019).

20. C. Tian, F. Pan, S. Xu, et al. Tunable magnetic properties in van der Waals crystals $(Fe_{1-x}Co_x)_5GeTe_2$. *Appl. Phys. Lett.* **116**, 202402 (2020).

21. R. Fujita, P. Bassirian, Z. Li, et al. Layer-Dependent Magnetic Domains in Atomically Thin $Fe_5GeTe_2$. *ACS Nano* **16**, 10545-10553 (2022).

22. C. Tan, W.-Q. Xie, G. Zheng, et al. Gate-Controlled Magnetic Phase Transition in a van der Waals Magnet $Fe_5GeTe_2$. *Nano Lett.* **21**, 5599-5605 (2021).

23. M. Ribeiro, G. Gentile, A. Marty, et al. Large-scale epitaxy of two-dimensional van der Waals room-temperature ferromagnet $Fe_5GeTe_2$. *npj 2D Mater. Appl.* **6**, 10 (2022).

24. Y. Gao, Q. Yin, Q. Wang, et al. Spontaneous (anti) Meron chains in the domain walls of van der Waals ferromagnetic $Fe_{5-x}GeTe_2$. *Adv. Mater.* **32**, 2005228 (2020).

25. H. Zhang, R. Chen, K. Zhai, et al. Itinerant ferromagnetism in van der Waals $Fe_{5-x}GeTe_2$ crystals above room temperature. *Phys. Rev. B* **102**, 064417 (2020).

26. Y. Deng, Z. Xiang, B. Lei, et al. Layer-Number-Dependent Magnetism and Anomalous Hall Effect in van der Waals Ferromagnet $Fe_5GeTe_2$. *Nano Lett.* **22**, 9839-9846 (2022).

27. M. Tang, J. Huang, F. Qin, et al. Continuous manipulation of magnetic anisotropy in a van der Waals ferromagnet via electrical gating. *Nat. Electron.* **6**, 28–36 (2023).

28. Z. Li, S. Liu, J. Sun, et al. Low-pass filters based on van der Waals ferromagnets. *Nat. Electron.* **6**, 273–280 (2023).

29. X. Lv, K. Pei, C. Yang, et al. Controllable Topological Magnetic Transformations in the Thickness Tunable van der Waals Ferromagnet $Fe_5GeTe_2$. *ACS Nano* **16**, 19319-19327 (2022).

30. Y. Zhang, S. Wang, Y. Feng, et al. Antisymmetric Peaks Observed in the Hall Resistance of $Fe_5GeTe_2$ Ferromagnets: Implications for Spintronic Devices. *ACS Appl. Nano Mater.* **6**, 15734-15740 (2023).

31. H. Wu, L. Chen, P. Malinowski, et al. Reversible Non-Volatile Electronic Switching in a Near Room Temperature van der Waals Ferromagnet. arxiv: 2307.03154v1.

32. M. Högen, R. Fujita, A. K. C. Tan, et al. Imaging Nucleation and Propagation of Pinned Domains in Few-Layer $Fe_{5-x}GeTe_2$. *ACS Nano* **17**, 16879-16885 (2023).

33. H. Zhang, Y.-T. Shao, R. Chen, et al. A room temperature polar magnetic metal. *Phys. Rev. Mater.* **6**, 044403 (2022).

34. X. Chen, Y.-T. Shao, R. Chen, et al. Pervasive beyond Room-Temperature Ferromagnetism in a Doped van der Waals Magnet. *Phys. Rev. Lett.* **128**, 217203 (2022).

35. H. Zhang, D. Raftrey, Y.-T. Chan et al., Room-temperature skyrmion lattice in a layered magnet $(Fe_{0.5}Co_{0.5})_5GeTe_2$, *Sci. Adv.* **8**, eabm7103 (2022).

36. A. K. Gopi, A. K. Srivastava, A. K. Sharma, et al. Thickness-Tunable Zoology of Magnetic Spin Textures Observed in $Fe_5GeTe_2$. *ACS Nano* **18**, 5335-5343 (2024).

37. B. W. Casas, Y. Li, A. Moon, et al. Coexistence of Merons with Skyrmions in the Centrosymmetric Van Der Waals Ferromagnet $Fe_{5-x}GeTe_2$. *Adv. Mater.* **35**, 2212087 (2023).

38. X. Chen, W. Tian, Y. He, et al. Thermal cycling induced alteration of the stacking order and spin-flip in the room temperature van der Waals magnet $Fe_5GeTe_2$. *Phys. Rev. Mater.* **7**, 044411 (2023).

39. M. Schmitt, T. Denneulin, A. Kovács, et al. Skyrmionic spin structures in layered $Fe_5GeTe_2$ up to room temperature. *Commun. Phys.* **5**, 254 (2022).
40. Y. Gao, S. Yan, Q. Yin, et al. Manipulation of topological spin configuration via tailoring thickness in van der Waals ferromagnetic $Fe_{5-x}GeTe_2$. *Phys. Rev. B* **105**, 014426 (2022).
41. K. Yamagami, Y. Fujisawa, M. Pardo-Almanza, et al. Enhanced d-p hybridization intertwined with anomalous ground state formation in the van der Waals itinerant magnet $Fe_5GeTe_2$. *Phys. Rev. B* **106**, 045137 (2022).
42. S. Ershadrad, S. Ghosh, D. Wang, et al. Unusual Magnetic Features in Two-Dimensional $Fe_5GeTe_2$ Induced by Structural Reconstructions. *J. Phys. Chem. Lett.* **13**, 4877-4883 (2022).
43. X. Wu, L. Lei, Q. Yin, et al. Direct observation of competition between charge order and itinerant ferromagnetism in the van der Waals crystal $Fe_{5-x}GeTe_2$. *Phys. Rev. B* **104**, 165101 (2021).
44. T. T. Ly, J. Park, K. Kim, et al. Direct observation of Fe-Ge ordering in $Fe_{5-x}GeTe_2$ crystals and resultant helimagnetism. *Adv. Funct. Mater.* **31**, 2009758 (2021).
45. M. Grelier, F. Godel, A. Vecchiola, et al. Three-dimensional skyrmionic cocoons in magnetic multilayers. *Nat. Commun.* **13**, 6843 (2022).
46. N. Nagaosa & Y. Tokura, Topological properties and dynamics of magnetic skyrmions. *Nat. Nanotech.* **8**, 899–911 (2013).
47. J. Lucassen, M. J. Meijer, O. Kurnosikov, et al. Tuning Magnetic Chirality by Dipolar Interactions. *Phys. Rev. Lett.* **123**, 157201 (2019).
48. T. Yokota, Numerical investigation of magnetic bubble types in a twodimensional ferromagnetic system with dipole–dipole interactions. *J. Phys. Soc. Jpn.* **88**, 084702 (2019).